\documentclass[aps,showpacs,preprintnumbers,amsmath,amssymb]
{revtex4}

\usepackage{graphicx}
\usepackage{dcolumn}
\usepackage{bm}
\usepackage{endnotes}
\begin{document}

\title{Vacuum fluctuation effects on asymmetric nuclear matter}
\author{ X.-H. Guo$^{1,}$\footnote[1]{ \emph{E-mail address:} xhguo@bnu.edu.cn},
 B. Liu$^{2,}$\footnote[2]{ \emph{E-mail address:}
 liub@mail.ihep.ac.cn},
 and M.-H. Weng$^{1,}$\footnote[3]{ Corresponding author, \emph{E-mail address:} mhweng@mail.bnu.edu.cn}}
\affiliation{$^{1}$ College of Nuclear Science and Technology, Beijing Normal University, Beijing 100875, China\\
$^{2}$ Institute of High Energy Physics, Chinese Academy of
Sciences, Beijing 100049, China}


\begin{abstract}
The vacuum fluctuation (VF) effects on asymmetric nuclear matter are
investigated. Masses of nucleons and mesons are modified in the
nuclear medium by calculating the loop-diagram corrections and the
density dependence of hadron masses is obtained. The relativistic
Lagrangian density with the isovector scalar $\delta$ meson is used
to calculate the nuclear equation of state (EOS) in the framework of
the relativistic mean-field (RMF) approach, the effects of the
in-medium hadron masses on the properties of neutron stars are
finally studied. With the inclusion of the VF corrections, the
nuclear EOS becomes softer and the neutron star masses are reduced.
\end{abstract}



\maketitle

\section*{I. Introduction}
\newenvironment
The The relativistic mean-field theory (RMF) \cite{walecka,brian},
which is one of the main applications of quantum hadrodynamics
(QHD),  is successful in nuclear structure studies
\cite{sugahara,lalazissis,sharma}. The nonlinear (NL) Walecka model
(NLWM), based on the RMF approach, has been extensively used to
study the properties of nuclear and neutron matter, $\beta$-stable
nuclei, and then extended to the drip-line regions
\cite{boguta,bodmer,mueller95,furnstahl,mueller96,liu02,liu05}. In
recent years some authors \cite{liu02,menezes,gaitanos,baran,liu05}
have stressed the importance of including the scalar  isovector
virtual $\delta(a_{0}(980))$ field in hadronic effective field
theories when asymmetric nuclear matter is studied. The inclusion of
the $\delta$ meson leads to the structure of relativistic
interactions, where a balance between an attractive (scalar) and a
repulsive (vector) potential exists. The $\delta$ meson plays the
role in the isospin channel and mainly affects the behavior of the
system at high density regions and so is of great interest in
nuclear astrophysics.

The properties of nuclear matter at high density play a crucial role
for building models of neutron stars (NS).
 Neutron stars are objects of highly compressed matter.
The structure of a compact star is characterized by its mass and
radius, which are obtained from appropriate equation of state (EOS)
at high densities. The EOS can be derived either from relativistic
or potential models.

In order to describe the medium dependence of nuclear interactions,
a density dependent relativistic hadron (DDRH) field theory has been
recently proposed \cite{liu0702,fuchs,jons,type}. Recently the
authors \cite{liu0702} used the density dependent coupling models
with the $\delta$ meson being included
 to study the neutron stars. They found that the introduction of the $\delta$ meson
 in the constant coupling model leads to heavier neutron stars in a nucleon-lepton picture.
The neutron star masses in the density dependent models can be
reduced when the $\delta$ meson is taken into account.

 The in-medium modification to the masses of $\sigma$, $\omega$,
 and $\rho$ mesons has been studied in experiments and
 theoretical approaches for a decade
\cite{brown,hatsuda,sarkar,ozawa,trnka,krusche,nasseripour}.
Recently the authors of Ref. \cite{abhijit} investigated the effect
of in-medium meson masses on the properties of the nuclear matter in
the Walecka model and the effective masses of  $\sigma$ and $\omega$
mesons in the nuclear medium were calculated by taking into account
the effects of the vacuum fluctuation (VF). In this work we want to
see the VF effects on asymmetric matter and neutron stars. We also
want to clarify the density dependence of in-medium nucleon and
meson masses. The VF effects are naturally introduced
 by considering loop corrections to the self-energies
of in-medium nucleons and mesons. The effective masses of nucleons
and mesons ($\sigma$, $\omega$, $\rho$, and $\delta$) in the nuclear
medium will be calculated in the VF-RMF model.
 The VF effects on asymmetric matter and
neutron stars will be studied.

This paper is organized as follows. In Sec. II, we derive the
in-medium effective masses of nucleons and mesons and the EOS for
nuclear matter in VF-RMF model. In Sec. III, we compare our results
with those of the NL-RMF model. In Sec. IV, a brief summary is
presented.

\section*{II. Hadron effective masses and EOS of nuclear matter}

The relativistic Lagrangian density of the interacting
many-particle system consisting of nucleons, isoscalar (scalar
$\sigma$, vector $\omega$), and isovector (scalar $\delta$, vector
$\rho$) mesons  used in this work is

\begin{widetext}
\begin{eqnarray}\label{eq:1}
{\cal
L}&=&\bar{\psi}\Bigl[i\gamma^{\mu}\partial_{\mu}-g_{\omega}\gamma^{\mu}\omega_{\mu}
-g_\rho\gamma^{\mu}\vec{\tau}\cdot\vec{b}_{\mu}-(M-g_{\sigma}\phi
-g_{\delta}\vec{\tau}\cdot\vec{\delta})\Bigr]\psi \nonumber\\
&&+\frac{1}{2}(\partial_{\mu}\phi\partial^{\mu}\phi-m_{\sigma}^{2}\phi^2)-U(\phi)
+\frac{1}{2}m_{\omega}^{2}\omega_{\mu}\omega^{\mu} \nonumber\\
&&+\frac{1}{2}m_{\rho}^{2}\vec{b}_{\mu}\cdot\vec{b}^{\mu}
+\frac{1}{2}(\partial_{\mu}\vec{\delta}\partial^{\mu}\vec{\delta}
-m_{\delta}^{2}\vec{\delta}^2) \nonumber\\
&&-\frac{1}{4}F_{\mu\nu}F^{\mu\nu}-\frac{1}{4}\vec{G}_{\mu\nu}\vec{G}^{\mu\nu}
 + \delta{\cal L},
\end{eqnarray}
\end{widetext}
where  $\phi$, $\omega_{\mu}$, $\vec{b}_{\mu}$, and $\vec{\delta}$
represent $\sigma$, $\omega$, $\rho$, and $\delta$ meson fields,
respectively, $U(\phi)=\frac{1}{3}a\phi^{3}+\frac{1}{4}b\phi^{4}$ is
 the nonlinear potential of the $\sigma$ meson,
$F_{\mu\nu}\equiv\partial_{\mu}\omega_{\nu}-\partial_{\nu}\omega_{\mu}$,
and
$\vec{G}_{\mu\nu}\equiv\partial_{\mu}\vec{b}_{\nu}-\partial_{\nu}\vec{b}_{\mu}$.
In order to remove the divergence in the loop calculation, the
counterterm for the Lagrangian density,
  $\delta{\cal L}$, included in Eq. (1) reads

\begin{eqnarray}\label{eq:2}
\delta{\cal
L}&=&\alpha_{1}\phi+\frac{1}{2!}\alpha_{2}\phi^{2}+\frac{1}{3!}\alpha_{3}\phi^{3}
+\frac{1}{4!}\alpha_{4}\phi^{4}+\frac{\zeta_{\sigma}}{2}(\partial\phi)^2\nonumber\\
&&+\beta_{1}\vec{\delta}+\frac{1}{2!}\beta_{2}\vec{\delta}^{2}+\frac{1}{3!}\beta_{3}\vec{\delta}^{3}
+\frac{1}{4!}\beta_{4}\vec{\delta}^{4}+\frac{\zeta_{\delta}}{2}(\partial\vec{\delta})^2\nonumber\\
&&+\frac{\zeta_{\omega}}{2}(\partial\omega_{\mu})^2+\frac{\zeta_{\rho}}{2}(\partial
\vec{b}_{\mu})^2,
\end{eqnarray}
where the parameters $\alpha_{i}$, $\beta_{i}$, and $\zeta_{j}$
($i$=1, 2, 3, 4; $j$=$\sigma$, $\omega$, $\rho$, $\delta$) are fixed
by the renormalization methods suggested in Refs. \cite{brian,
haruki}.

The field equations in the RMF approximation are

\begin{eqnarray}\label{eq:3}
[i\gamma^{\mu}\partial_{\mu}-(M- g_{\sigma}\phi
-g_\delta{\tau_3}\delta_3)-g_\omega\gamma^{0}{\omega_0}-g_\rho\gamma^{0}{\tau_3}{b_0}]\psi&=&0,\label{eq:6-1}\nonumber\\
m_{\sigma}^{2}\phi+a\phi^2+b\phi^3&=&g_{\sigma}\rho^{s},\label{eq:6-2}\nonumber\\
m_{\omega}^{2}\omega_{0}&=&g_{\omega}\rho,\label{eq:6-3}\nonumber\\
m_{\rho}^{2} b_{0}&=&g_{\rho}\rho_{3},\label{eq:6-4}\nonumber\\
m_{\delta}^{2} \delta_{3}&=&g_{\delta}\rho_{3}^{s},
\end{eqnarray}
where $\rho^{(s)}=\rho_{p}^{(s)}+\rho_{n}^{(s)}$ and
$\rho_{3}^{(s)}=\rho_{p}^{(s)}-\rho_{n}^{(s)}$, where $\rho_{i}$ and
$\rho^{s}_{i}$ (the index (subscript or superscript) $i$ denotes
proton or neutron throughout this paper) are the nucleon and scalar
densities, respectively.

The nucleon and the scalar densities are given by, respectively,

\begin{eqnarray}\label{eq:4}
\rho_{i}&=&\frac{k_{F_{i}}^{3}}{3\pi^2},
\end{eqnarray}
and
\begin{eqnarray}\label{eq:5}
\rho_{i}^{s}&=& - i\int \frac{{\rm d} ^4k}{(2\pi)^4}{\rm
Tr}G^{i}(k),
\end{eqnarray}
where $k_{F_{i}}$ is the Fermi momentum of the nucleon  and
$G^{i}(k)$ is the nucleon propagator in the VF-RMF model:

\begin{eqnarray}\label{eq:6}
G^{i}(k)&=&(\gamma_{\mu}k^{\mu}+M_{i}^{\star})\Bigl[\frac{1}{k^2-M_{i}^{\star2}+i\eta}
+\frac{i\pi}{E_{F_{i}}^{\star}}\delta(k^0-E_{F_{i}})\theta(k_{F_{i}}-|\vec{k}|)\Bigr]\nonumber\\
&\equiv&G^{i}_{F}(k)+G^{i}_{D}(k),
\end{eqnarray}
where $E_{F_{i}}^{\star}=\sqrt{k_{F_{i}}^2+M_{i}^{\star2}}$,
$M_{i}^{\star}$ is the nucleon effective masses, and $\eta$ is
infinitesimal. Here $G^{i}_{F}(k)$ describes the free propagation of
positive and negative energy quasi-nucleons. $G^{i}_{D}(k)$
describes quasi-nucleon 'holes' inside the Fermi sea and corrects
the propagation of positive energy quasi-nucleons by the Pauli
exclusion principle.

\begin{figure}[hbtp]
\begin{center}
\includegraphics[scale=0.8]{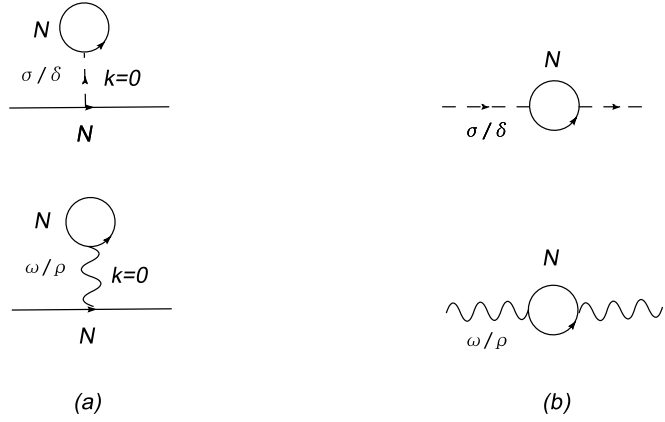}
\caption{Loop-diagram corrections to the self-energy of nucleons (a)
and mesons (b) in nuclear medium, where $N$ denotes nucleon and
 $k$ is the four momentum of the meson.}\label{fig01}
\end{center}
\end{figure}

The nucleon effective mass without the $\delta$ field in the RMF
approach is $M^\star=M-g_\sigma \phi$. When the VF effects are
considered, the loop-diagram corrections to the self-energy of
nucleons and mesons shown in Fig. 1 are naturally introduced. The
nucleon effective mass without the $\delta$ field in the RMF
approach including the vacuum fluctuations can be  calculated from
Fig. 1(a). Thus, we have

\begin{eqnarray}\label{eq:7}
M^{\star}&=&M + \frac{ig_{\sigma} ^{2}}{m_{\sigma}^{\star 2}}\int
\frac{{\rm d} ^4k}{(2\pi)^4}[{\rm
 Tr}G^{p}(k)+{\rm
 Tr}G^{n}(k)]
+\frac{a}{m_{\sigma}^{\star2}g_{\sigma}}(M^{\star}-M)^2
-\frac{b}{m_{\sigma}^{\star2}g_{\sigma}^{2}}(M^{\star}-M)^3\nonumber\\
&=&M-\frac{g_{\sigma}^{2}}{2\pi^2m_{\sigma}^{\star2}}
\Bigl[k_{F_{p}}M^{\star}E_{F_{p}}^{\star}-M^{\star3}{\rm
ln}(\frac{k_{F_{p}}+E_{F_{p}}^{\star}}{M^{\star}})\Bigr] \nonumber\\
&&-\frac{g_{\sigma}^{2}}{2\pi^2m_{\sigma}^{\star2}}
\Bigl[k_{F_{n}}M^{\star}E_{F_{n}}^{\star}-M^{\star3}{\rm ln}(\frac{k_{F_{n}}+E_{F_{n}}^{\star}}{M^{\star}})\Bigr] \nonumber\\
  &&+\frac{g_{\sigma}^{2}}{\pi^2m_{\sigma}^{\star2}}
\Bigl[M^{\star3}\ln(\frac{M^{\star}}{M})-M^2(M^{\star}-M)-\frac{5}{2}M(M^{\star}-M)^2-\frac{11}{6}(M^{\star}-M)^3\Bigr]\nonumber\\
  &&+\frac{a}{m_{\sigma}^{\star2}g_{\sigma}}(M^{\star}-M)^2
-\frac{b}{m_{\sigma}^{\star2}g_{\sigma}^{2}}(M^{\star}-M)^3,
\end{eqnarray}
where the second term in the right hand side of the first line in
Eq. (7) is given by Fig. 1(a),  and the third and fourth terms are
the contributions from the nonlinear potential of the $\sigma$
meson.

 As well known, the $\delta$ meson leads to a definite splitting of proton and
 neutron effective masses,
 the nucleon effective masses with the $\delta$ meson in the RMF approach are given by, respectively,

\begin{eqnarray}\label{eq:8}
M_p^{\star}&=&M - g_{\sigma} \phi -g_{\delta} \delta_3,
\end{eqnarray}
and
\begin{eqnarray}\label{eq:9}
M_n^{\star}&=&M- g_{\sigma} \phi + g_{\delta} \delta_3.
\end{eqnarray}

The nucleon effective masses with the $\delta$ meson in the RMF
approach including the vacuum fluctuations can be calculated from
Fig. 1(a),

\begin{eqnarray}\label{eq:10}
 M_{p}^{\star}&=&M - g_{\sigma} \phi-g_{\delta} \delta_3\nonumber\\
 &=&M + \frac{ig_{\sigma} ^{2}}{m_{\sigma}^{\star 2}}\int \frac{{\rm d} ^4k}{(2\pi)^4}[{\rm
 Tr}G^{p}(k)+{\rm Tr}G^{n}(k)]\nonumber\\
 && + \frac{ig_{\delta} ^{2}}{m_{\delta}^{\star 2}}\int \frac{{\rm d} ^4k}{(2\pi)^4}[{\rm
 Tr}G^{p}(k)-{\rm
 Tr}G^{n}(k)]\nonumber\\
 &&-a\frac{2g_{\sigma}^{2}}{m_{\sigma}^{\star2}}\frac{(M_{p}^{\star}+M_{n}^{\star}-2M)^2}{(-2g_{\sigma})^3}
 -b\frac{2g_{\sigma}^{2}}{m_{\sigma}^{\star2}}\frac{(M_{p}^{\star}+M_{n}^{\star}-2M)^3}{(-2g_{\sigma})^4},
\end{eqnarray}
and
\begin{eqnarray}\label{eq:11}
 M_{n}^{\star}&=&M - g_{\sigma} \phi+ g_{\delta} \delta_3\nonumber\\
 &=&M + \frac{ig_{\sigma} ^{2}}{m_{\sigma}^{\star 2}}\int \frac{{\rm d} ^4k}{(2\pi)^4}[{\rm
 Tr}G^{n}(k)+{\rm
 Tr}G^{p}(k)]\nonumber\\
 &&+ \frac{ig_{\delta} ^{2}}{m_{\delta}^{\star 2}}\int \frac{{\rm d} ^4k}{(2\pi)^4}[{\rm
 Tr}G^{n}(k)-{\rm
 Tr}G^{p}(k)]\nonumber\\
 &&-a\frac{2g_{\sigma}^{2}}{m_{\sigma}^{\star2}}\frac{(M_{p}^{\star}+M_{n}^{\star}-2M)^2}{(-2g_{\sigma})^3}
 -b\frac{2g_{\sigma}^{2}}{m_{\sigma}^{\star2}}\frac{(M_{p}^{\star}+M_{n}^{\star}-2M)^3}{(-2g_{\sigma})^4}.
\end{eqnarray}

 Thus, we get after the momentum integral
\begin{eqnarray}\label{eq:12}
   M^{\star}_{p}&=&M-\frac{1}{2\pi^2}(\frac{g_{\sigma}^{2}}{m_{\sigma}^{\star2}}
  +\frac{g_{\delta}^{2}}{m_{\delta}^{\star2}})
  \Bigl[k_{F_{p}}M^{\star}_{p}E_{F_{p}}^{\star}-M_{p}^{\star3}{\rm ln}(\frac{k_{F_{p}}+E_{F_{p}}^{\star}}{M_{p}^{\star}})\Bigr] \nonumber\\
  &&+\frac{1}{2\pi^2}(\frac{g_{\sigma}^{2}}{m_{\sigma}^{\star2}}
  +\frac{g_{\delta}^{2}}{m_{\delta}^{\star2}})
 \Bigl[M^{\star3}_{p}\ln(\frac{M^{\star}_{p}}{M})-M^2(M^{\star}_{p}-M)\nonumber\\
  &&-\frac{5}{2}M(M^{\star}_{p}-M)^2-\frac{11}{6}(M^{\star}_{p}-M)^3\Bigr]\nonumber\\
  &&-\frac{1}{2\pi^2}(\frac{g_{\sigma}^{2}}{m_{\sigma}^{\star2}}
  -\frac{g_{\delta}^{2}}{m_{\delta}^{\star2}})
 \Bigl[k_{F_{n}}M^{\star}_{n}E_{F_{n}}^{\star}-M_{n}^{\star3}{\rm ln}(\frac{k_{F_{n}}+E_{F_{n}}^{\star}}{M_{n}^{\star}})\Bigr] \nonumber\\
  &&+\frac{1}{2\pi^2}(\frac{g_{\sigma}^{2}}{m_{\sigma}^{\star2}}
  -\frac{g_{\delta}^{2}}{m_{\delta}^{\star2}})
  \Bigl[M^{\star3}_{n}\ln(\frac{M^{\star}_{n}}{M})-M^2(M^{\star}_{n}-M)\nonumber\\
  &&-\frac{5}{2}M(M^{\star}_{n}-M)^2-\frac{11}{6}(M^{\star}_{n}-M)^3\Bigr]\nonumber\\
  &&-a\frac{2g_{\sigma}^{2}}{m_{\sigma}^{\star2}}\frac{(M_{p}^{\star}+M_{n}^{\star}-2M)^2}{(-2g_{\sigma})^3}
-b\frac{2g_{\sigma}^{2}}{m_{\sigma}^{\star2}}\frac{(M_{p}^{\star}+M_{n}^{\star}-2M)^3}{(-2g_{\sigma})^4},
\end{eqnarray}
and
\begin{eqnarray}\label{eq:13}
  M^{\star}_{n}&=&M-\frac{1}{2\pi^2}(\frac{g_{\sigma}^{2}}{m_{\sigma}^{\star2}}
  -\frac{g_{\delta}^{2}}{m_{\delta}^{\star2}})
  \Bigl[k_{F_{p}}M^{\star}_{p}E_{F_{p}}^{\star}-M_{p}^{\star3}{\rm ln}(\frac{k_{F_{p}}+E_{F_{p}}^{\star}}{M_{p}^{\star}})\Bigr] \nonumber\\
  &&+\frac{1}{2\pi^2}(\frac{g_{\sigma}^{2}}{m_{\sigma}^{\star2}}
  -\frac{g_{\delta}^{2}}{m_{\delta}^{\star2}})
  \Bigl[M^{\star3}_{p}\ln(\frac{M^{\star}_{p}}{M})-M^2(M^{\star}_{p}-M)\nonumber\\
  &&-\frac{5}{2}M(M^{\star}_{p}-M)^2-\frac{11}{6}(M^{\star}_{p}-M)^3\Bigr]\nonumber\\
  &&-\frac{1}{2\pi^2}(\frac{g_{\sigma}^{2}}{m_{\sigma}^{\star2}}
  +\frac{g_{\delta}^{2}}{m_{\delta}^{\star2}})
  \Bigl[k_{F_{n}}M^{\star}_{n}E_{F_{n}}^{\star}-M_{n}^{\star3}{\rm ln}(\frac{k_{F_{n}}+E_{F_{n}}^{\star}}{M_{n}^{\star}})\Bigr] \nonumber\\
  &&+\frac{1}{2\pi^2}(\frac{g_{\sigma}^{2}}{m_{\sigma}^{\star2}}
  +\frac{g_{\delta}^{2}}{m_{\delta}^{\star2}})
  \Bigl[M^{\star3}_{n}\ln(\frac{M^{\star}_{n}}{M})-M^2(M^{\star}_{n}-M)\nonumber\\
  &&-\frac{5}{2}M(M^{\star}_{n}-M)^2-\frac{11}{6}(M^{\star}_{n}-M)^3\Bigr]\nonumber\\
  &&-a\frac{2g_{\sigma}^{2}}{m_{\sigma}^{\star2}}\frac{(M_{p}^{\star}+M_{n}^{\star}-2M)^2}{(-2g_{\sigma})^3}
-b\frac{2g_{\sigma}^{2}}{m_{\sigma}^{\star2}}\frac{(M_{p}^{\star}+M_{n}^{\star}-2M)^3}{(-2g_{\sigma})^4},
\end{eqnarray}
where $m_{j}^{\star} ~(j=\sigma, \omega, \rho, \delta)$ are the
in-medium meson masses.

The effective mass of the meson (or in-medium meson mass) is defined
as the pole position of the meson propagator at zero three-momentum
(on-shell) or zero four-momentum (off-shell). We calculate the
in-medium meson masses in the random phase approximation (RPA)
\cite{abhijit, haruki}. The corresponding diagrams are given in Fig.
1(b). Thus, we obtain the in-medium masses of scalar mesons:

\begin{equation}\label{eq:14}
 m_{j}^{\star2}=m_{j}^{2}+\Pi_{j}(q^{\mu})\ \, (j=\sigma,\delta),
\end{equation}
where
\begin{equation}\label{eq:15}
\Pi_{j}(q^{\mu})=-ig_{j}^{2}\sum\limits_{i=n,p}\int\frac{{\rm
d^4}k}{(2\pi)^4}{\rm Tr} G^{i}(k+q)G^{i}(k).
\end{equation}

The expressions of $\Pi_{j}$ (on-shell and off-shell) are, respectively,

\begin{eqnarray}\label{eq:16}
\Pi_{j}(\vec{q}=0;
q^{0}=m_{j}^{\star})&=&\frac{g_{j}^{2}}{4\pi^2}\sum\limits_{i=n,p}\Bigl[\frac{1}{m_{j}^{\star}}(4M_{i}^{\star2}-m_{j}^{\star2})^{3/2}
\arctan(\frac{m_{j}^{\star}k_{F_{i}}}{E_{F_{i}}^{\star}\sqrt{ 4M_{i}^{\star2}-m_{j}^{\star2}}})\nonumber\\
&&+(m_{j}^{\star2}-6M_{i}^{\star2}){\rm ln}\Bigl(\frac{k_{F_{i}}+E_{F_{i}}^{\star}}{M_{i}^{\star}}\Bigr)+2k_{F_{i}}E_{F_{i}}^{\star}\Bigr]\nonumber\\
&&-\frac{g^{2}_{j}}{2\pi^2}\sum\limits_{i=n,p}\biggl\{\frac{3}{2}(M_{i}^{\star2}-M^{2})\int\limits_{0}^{1}
{\rm d}x \ln\Bigl(1-\frac{m_{j}^{2}}{M^{2}}x(1-x)\Bigr)\nonumber\\
&&+\frac{3}{2}\int\limits_{0}^{1}{\rm
d}x\Bigl[(M_{i}^{\star2}-m_{j}^{\star2}x(1-x))
\ln\Bigl(\frac{M_{i}^{\star2}-m_{j}^{\star2}x(1-x)}{M^{2}-m_{j}^{2}x(1-x)}\Bigr)\Bigr]\nonumber\\
&&-\frac{m_{j}^{2}-m_{j}^{\star2}}{4}-3M(M_{i}^{\star}-M)
-\frac{9}{2}(M_{i}^{\star}-M)^2\biggr\},
\end{eqnarray}
and
\begin{eqnarray}\label{eq:17}
\Pi_{j}(q^{\mu}=0)&=&\frac{g_{j}^{2}}{2\pi^2}\sum\limits_{i=n,p}\Bigl[\frac{3M_{i}^{\star2}k_{F_{i}}+k_{F_{i}}^{3}}{E_{F_{i}}^{\star}}
-3M_{i}^{\star2}{\rm ln}\Bigl(\frac{k_{F_{i}}+E_{F_{i}}^{\star}}{M_{i}^{\star}}\Bigr)\Bigr]\nonumber\\
&&-\frac{3g_{j}^{2}}{4\pi^2}\sum\limits_{i=n,p}\Bigl[2M_{i}^{\star2}
\ln\Bigl(\frac{M_{i}^{\star}}{M}\Bigr)-M^{2}+4MM_{i}^{\star}-3M_{i}^{\star2}\Bigr].
\end{eqnarray}

The effective masses of vector mesons can be obtained from Fig. 1(b)

\begin{equation}\label{eq:18}
 m_{j}^{\star2}=m_{j}^{2}+\Pi_{j T}(q^{\mu})\ \ (j=\omega,\rho),
\end{equation}
where $\Pi_{j T}$ is the transverse part of the polarization tensor
as follows:

\begin{equation}\label{eq:19}
\Pi_{\mu\nu}(q^{\mu})=-ig_{j}^{2}\sum\limits_{i=n,p}\int\frac{{\rm
 d^4}k}{(2\pi)^4}{\rm Tr}
 \gamma_{\mu}G^{i}(k+q)\gamma_{\nu}G^{i}(k).
\end{equation}

So the expressions of $\Pi_{j T}$ (on-shell and off-shell) are, respectively,

\begin{eqnarray}\label{eq:20}
\Pi_{j T}(\vec{q}=0;
q^{0}=m_{j}^{\star})&=&-\frac{g_{j}^{2}}{6\pi^2}\sum\limits_{i=n,p}\Bigl[
\frac{8M_{i}^{\star4}+2M_{i}^{\star2}m_{j}^{\star2}-m_{j}^{\star4}}{m_{j}^{\star}\sqrt{4M_{i}^{\star2}-m_{j}^{\star2}}}
\arctan\Bigl(\frac{m_{j}^{\star}k_{F_{i}}}{E_{F_{i}}^{\star}\sqrt{4M_{i}^{\star2}-m_{j}^{\star2}}}\Bigr)\nonumber\\
&&-2k_{F_{i}}E_{F_{i}}^{\star}-m_{j}^{\star2}{\rm ln}\Bigl(\frac{k_{F_{i}}+E_{F_{i}}^{\star}}{M_{i}^{\star}}\Bigr)\Bigr]\nonumber\\
&&-\frac{g_{j}^{2}m_{j}^{\star2}}{2\pi^2}\sum\limits_{i=n,p}\int\limits_{0}^{1}{\rm
d}xx(x-1)
\ln\Bigl[\frac{M_{i}^{\star2}-m_{j}^{\star2}x(1-x)}{M^{2}-m_{j}^{2}x(1-x)}\Bigr],
\end{eqnarray}
and

\begin{equation}\label{eq:21}
\Pi_{j
T}(q^{\mu}=0)=\frac{g_{j}^{2}}{3\pi^2}\sum\limits_{i=n,p}\frac{k_{F_{i}}^{3}}{E_{F_{i}}^{\star}}.
\end{equation}

We note that the VF contributions are included in the second summations
of $\Pi_{j}(q^{\mu})$ and $\Pi_{jT}(q^{\mu})$ (Eqs. (16), (17) and (20)),
 the VF contribution in Eq. (21) equals zero.

The in-medium meson propagator is significantly modified by the
interaction of the mesons with the nucleons. Clearly, this
modification of the meson propagators will change the nucleon
effective mass as well as the energy density. The meson propagators
in the tadpole diagram are calculated at zero four momentum
transfer. So we must replace the meson mass appearing in the nucleon
effective mass and the energy density by the meson effective mass
defined as \cite{abhijit}:

\begin{equation}\label{eq:22}
 m_{j}^{\star2}=m_{j}^{2}+\Pi_{j}(q^{\mu}=0)\ \, (j=\sigma,\delta),
\end{equation}
and
\begin{equation}\label{eq:23}
 m_{j}^{\star2}=m_{j}^{2}+\Pi_{j T}(q^{\mu}=0)\ \ (j=\omega,\rho).
\end{equation}

Therefore, the energy-momentum tensor in the VF-RMF model can be expressed as

\begin{eqnarray}\label{eq:24}
T_{\mu\nu}&=&i{\bar
\psi}\gamma_{\mu}\partial_{\nu}\psi+g_{\mu\nu}\biggl[\frac{1}{2}m_{\sigma}^{\star}\phi^{2}
+\frac{1}{2}m_{\delta}^{\star}{\vec
\delta}^{2}-\frac{1}{2}m_{\omega}^{\star}\omega_{\lambda}\omega^{\lambda}
-\frac{1}{2}m_{\rho}^{\star}{\vec b_{\lambda}}{\vec
b^{\lambda}}+U(\phi)\biggr].
\end{eqnarray}

Thus the EOS for nuclear matter in the VF-RMF model can be obtained.  The energy density is given by

\begin{eqnarray}\label{eq:25}
  \varepsilon&=&\langle T^{00}\rangle=\frac{g_{\omega}^2}{2m_{\omega}^{\star2}}\rho^2
  +\frac{g_{\rho}^2}{2m_{\rho}^{\star2}}\rho_{3}^2
  +\frac{1}{2}m_{\sigma}^{\star2}\phi^2
  +\frac{1}{2}m_{\delta}^{\star2}\delta_{3}^2\nonumber\\
  &&+\frac{1}{8\pi^2}\sum\limits_{i=n,p}\Bigl[k_{F_{i}}E_{F_{i}}^{\star}(M_{i}^{\star2}+2k_{F_{i}}^{2})
  -M_{i}^{\star4}{\rm ln}(\frac{k_{F_{i}}+E_{F_{i}}^{\star}}{M_{i}^{\star}}) \Bigr]+U(\phi) \nonumber\\
  &&-\frac{1}{8\pi^2}\sum\limits_{i=n,p}\Bigl[M_{i}^{\star4}{\rm ln}(\frac{M_{i}^{\star}}{M})+M^{3}(M-M_{i}^{\star})
  -\frac{7}{2}M^2(M-M_{i}^{\star})^2\nonumber\\
  &&+\frac{13}{3}M(M-M_{i}^{\star})^3-\frac{25}{12}(M-M_{i}^{\star})^4\Bigr],
\end{eqnarray}
and the pressure is

\begin{eqnarray}\label{eq:26}
  P&=&\frac{1}{3}\langle T^{ii}\rangle=\frac{g_{\omega}^2}{2m_{\omega}^{\star2}}\rho^2
  +\frac{g_{\rho}^2}{2m_{\rho}^{\star2}}\rho_{3}^2
  -\frac{1}{2}m_{\sigma}^{\star2}\phi^2
  -\frac{1}{2}m_{\delta}^{\star2}\delta_{3}^2\nonumber\\
  &&+\frac{1}{8\pi^2}\sum\limits_{i=n,p}\Bigl[M_{i}^{\star4}{\rm ln}(\frac{k_{F_{i}}
  +E_{F_{i}}^{\star}}{M_{i}^{\star}})-E_{F_{i}}^{\star}k_{F_{i}}(M_{i}^{\star2}-\frac{2}{3}k_{F_{i}}^{2})\Bigr]-U(\phi)\nonumber\\
  &&+\frac{1}{8\pi^2}\sum\limits_{i=n,p}\Bigl[M_{i}^{\star4}{\rm ln}(\frac{M_{i}^{\star}}{M})+M^{3}(M-M_{i}^{\star})
  -\frac{7}{2}M^2(M-M_{i}^{\star})^2\nonumber\\
  &&+\frac{13}{3}M(M-M_{i}^{\star})^3-\frac{25}{12}(M-M_{i}^{\star})^4\Bigr].
\end{eqnarray}

The density dependence of the nuclear symmetry energy, $E_{sym}$, is
one of the basic properties of asymmetric nuclear matter for
studying the structure of neutron stars. Empirically, we have
information on $E_{sym}$ only at the saturation point, where it
ranges from 28 to 35 MeV according to the nuclear mass table
\cite{myers}. The nuclear symmetry energy is defined through the
expansion of the binding energy in terms of the asymmetry parameter
$\alpha$ \cite{liu05}:

\begin{equation}\label{eq:27}
E/A(\rho,\alpha)=E/A(\rho,0)+E_{sym}(\rho)\alpha^2+O(\alpha^4)+\cdot\cdot\cdot,
\end{equation}
where the binding energy density is defined as
$E/A=\varepsilon/\rho-M$, and the asymmetry parameter
$\alpha=(\rho_{n}-\rho_{p})/\rho$.

The nuclear symmetry energy is defined by

\begin{equation}\label{eq:28}
E_{sym}\equiv\frac{1}{2}\frac{\partial^{2}(E/A)}{\partial\alpha^{2}}\Big|_{\alpha=0}
=\frac{1}{2}\rho\frac{\partial^{2}\varepsilon}{\partial\rho^{2}_{3}}\Big|_{\rho_{3}=0}.
\end{equation}

According to the definition, an explicit expression for the symmetry energy in the VF-RMF model
is obtained as

\begin{eqnarray}\label{eq:29}
E_{sym}&=&\frac{1}{2}\frac{g_{\rho}^{2}}{m_{\rho}^{\star2}}\rho+\frac{1}{6}\frac{k_{F}^{2}}{E_{F}^{\star}}
-\frac{1}{2}\frac{g_{\delta}^{2}}{m_{\delta}^{\star2}}\frac{M^{\star2}\rho}{E_{F}^{\star2}
(1+\frac{g_{\delta}^{2}}{m_{\delta}^{\star2}}C-\frac{1}{\pi^{2}}\frac{g_{\delta}^{2}}{m_{\delta}^{\star2}}D)}\nonumber\\
&&+\frac{\rho}{2\pi^2}\frac{g_{\delta}^{4}}{m_{\delta}^{\star4}}\frac{M^{\star2}}{E_{F}^{\star2}}
\frac{D}{1+\frac{g_{\delta}^{2}}{m_{\delta}^{\star2}}C-\frac{1}{\pi^{2}}\frac{g_{\delta}^{2}}{m_{\delta}^{\star2}}D}\nonumber\\
&&-\frac{\rho}{4\pi^2}\frac{g_{\delta}^{4}}{m_{\delta}^{\star4}}
\frac{M^{\star2}}{E_{F}^{\star2}}\frac{1}{(1+\frac{g_{\delta}^{2}}{m_{\delta}^{\star2}}C-\frac{1}{\pi^{2}}
\frac{g_{\delta}^{2}}{m_{\delta}^{\star2}}D)^2}\Bigr[12M^{\star2}{\rm
ln}\frac{M^{\star}}{M}+7M^{\star2}\nonumber\\
&&-7M^{2}+26M(M-M^{\star})-25(M-M^{\star})^2\Bigr],
\end{eqnarray}
where
\begin{equation}
  C=\frac{1}{\pi^2}\Big[\frac{k_{F}E_{F}^{\star2}+2M^{\star2}k_{F}}{E_{F}^{\star}}
  -3M^{\star2}{\rm
  ln}\Bigl(\frac{k_{F}+E_{F}^{\star}}{M^{\star}}\Bigr)\Big],
\end{equation}
and
\begin{equation}\label{eq:30}
D=3M^{\star2}{\rm
ln}\frac{M^{\star}}{M}+M^{\star2}-M^{2}-5M(M^{\star}-M)-\frac{11}{2}(M^{\star}-M)^2.
\end{equation}

As discussed in Refs. \cite{norman,kouno}, the incompressibility $K$
is one of the important ingredients for the nuclear EOS. The
incompressibility of nuclear matter is defined by
\cite{norman,kouno}

\begin{equation}\label{eq:31}
K=9\rho_{0}^{2}\frac{\partial^{2}(\varepsilon/\rho)}{\partial
\rho^{2}}\Big|_{\rho=\rho_{0}}=9\frac{\partial P}{\partial
\rho}\Big|_{\rho=\rho_{0}}=9\rho_{0}\frac{\partial \mu}{\partial
\rho}\Big|_{\rho=\rho_{0}},
\end{equation}
where $\mu=(\varepsilon+P)/\rho$ is the baryon chemical potential
and $\rho_{0}$ is the saturation density.

We can easily obtain the incompressibility from the EOS in the VF-RMF model as

\begin{equation} \label{eq:32}
K=9\rho_{0}\Bigl(\frac{k_{F}^{2}}{3\rho
E_{F}^{\star}}+\frac{g_{\omega}^{2}}{m_{\omega}^{\star2}}+\frac{M^{\star}}{E_{F}^{\star}}\frac{\partial
M^{\star}}{\partial \rho}\Bigr)\Big|_{\rho=\rho_{0}},
\end{equation}
where

\begin{equation}\label{eq:33}
\frac{\partial M^{\star}}{\partial
\rho}=-\frac{g_{\sigma}^{2}}{m_{\sigma}^{\star2}}\frac{M^{\star}}{E_{F}^{\star}}Q^{-1},
\end{equation}
and
\begin{eqnarray}\label{eq:34}
Q&=&1+\frac{g_{\sigma}^{2}}{\pi^{2}m_{\sigma}^{\star2}}\Bigl(k_{F}E_{F}^{\star}+\frac{2k_{F}M^{\star2}}{E_{F}^{\star}}-3M^{\star2}{\rm
ln}(\frac{k_{F}+E_{F}^{\star}}{M})+\frac{9}{2}M^{\star2}+\frac{3}{2}M^{2}-6MM^{\star}\Bigr)\nonumber\\
&&+\frac{2a}{m_{\sigma}^{\star2}g_{\sigma}}(M-M^{\star})+\frac{3b}{m_{\sigma}^{\star2}g_{\sigma}^{2}}(M-M^{\star})^{2}.
\end{eqnarray}

The final outcome of a supernova explosion can be a neutron star or
a black hole. The neutron star is believed to evolve from an
initially hot protoneutron  star. The matter in cold neutron stars
is in the ground state in nuclear equilibrium. Matter in equilibrium
concerning weak interactions is called as $\beta$-equilibrium
matter. The composition of $\beta$-equilibrium system is determined
by the request of charge neutrality and chemical-potential
equilibrium \cite{norman,lattimer}. The balance processes for
$\beta$-equilibrium ($npe$) system are the following weak reactions:
\begin{eqnarray}\label{eq:35}
 n&\longrightarrow &p+e^{-}+\bar{\nu}_{e},\\
 p+e^{-}&\longrightarrow& n+\nu_{e}.
\end{eqnarray}

The chemical-potential equilibrium condition for ($npe$) system can
be written as

\begin{equation}\label{eq:36}
\mu_{e}=\mu_{n}-\mu_{p},
\end{equation}
where the  electron chemical-potential $\mu_{e}=\sqrt{k_{F_{e}}^{2}+m^{2}_{e}}$,
$k_{F_{e}}$ is the electron momentum at the fermion level and
$m_{e}$ is the electron mass.

The charge neutrality condition is
\begin{equation}\label{eq:37}
\rho_{e}=\rho_{p}=X_{p}\rho,
\end{equation}
where $\rho_{e}$ is the electron density, and the proton fraction
$X_{p}=Z/A=\rho_{p}/\rho$.

 In the ultra-relativistic limit for
 noninteracting electrons, the electron density can be expressed as a function of its
 chemical potential

 \begin{equation}\label{eq:38}
 \rho_{e}=\frac{1}{3\pi^{2}}\mu_{e}^{3}.
 \end{equation}

 Then, we can obtain the relation between the proton fraction $X_{p}$ and the nuclear symmetry
 energy $E_{sym}$

\begin{equation}\label{eq:39}
 3\pi^{2}\rho X_{p}-[4E_{sym}(1-2X_{p})]^{3}=0.
\end{equation}

The EOS for $\beta$-equilibrium ($npe$) matter can be estimated by
using the values of $X_{p}$,
 which can be obtained by solving Eq. (41).
 The properties of the neutron stars can be finally studied by
 solving Tolmann-Oppenheimer-Volkov (TOV) equations \cite{tolman} with the derived
 nuclear EOS as an input.

\section*{III. Results and discussions}

In order to make a comparison with the NL-RMF model
\cite{liu02,liu05}, the same saturation properties of nuclear matter
and hadron masses are listed in Table I, which are used to determine
the model parameters. The obtained model parameters are presented in
Table II with the NL-RMF model parameters together for a comparison.
The coupling constants are defined as $f_{j}=g_{j}^{2}/m_{j}^{2}$
($j=\sigma$, $\omega$, $\rho$, $\delta$) in Refs.
\cite{liu02,liu05}. The parameters of self-interacting terms in
Table II are defined as $A=a/g_{\sigma}^{3}$ and
$B=b/g_{\sigma}^{4}$.

\begin{center}
{{\large \bf Table I.}~Saturation properties of nuclear matter and hadron masses.}

\par
\vspace{0.5cm} \noindent

\begin{tabular}{ c c c } \hline
$saturation ~properties$ &\cite{liu05} \\ \hline
$\rho_{0}~(fm^{-3})$    &0.16     \\ \hline
$E/A ~(MeV)$            &-16.0    \\ \hline
$K~(MeV)$               &240.0    \\ \hline
$E_{sym}~(MeV)$         &31.3     \\ \hline
$M^{\star}/M $          &0.75     \\ \hline
$M~(MeV)$               &939      \\ \hline
$m_{\sigma}~(MeV)$      &550      \\ \hline
$m_{\omega}~(MeV)$      &783      \\ \hline
$m_{\rho}~(MeV)$        &770      \\ \hline
$m_{\delta}~(MeV)$      &980      \\ \hline
\end{tabular}
\end{center}

\par
\vspace{0.3cm} \noindent
{{\large \bf Table II.}~Model Parameters in
the VF-RMF and NL-RMF models.}
\par
\begin{center}
\vspace{0.5cm} \noindent
\begin{tabular}{c|c|c|c|c}  \hline
$Parameter$    &\multicolumn{2}{|c}{$VF-RMF$ model}
               &\multicolumn{2}{|c}{$NL-RMF$ model}\cite{liu05} \\ \cline{2-5}
               &$VF\rho$  &$VF\rho\delta$    &$NL\rho$   &$NL\rho\delta$ \\\hline
 $g_\sigma$    &12.33     &12.33             &8.96       &8.96           \\\hline
 $g_\omega$    &10.52     &10.52             &9.24       &9.24           \\\hline
 $g_\rho$      &4.01      &6.80              &3.80       &6.93           \\\hline
 $g_\delta$    &0.00      &7.85              &0.00       &7.85           \\\hline
 $A~(fm^{-1})$ &0.048     &0.048             &0.033      &0.033          \\\hline
 $B$           &-0.021    &-0.021            &-0.0048    &-0.0048        \\\hline
\end{tabular}
\end{center}
\vspace{0.5cm}

We use the obtained parameters in Table II to complete the
calculations of self-consistence in the present work. The masses of
hadrons (nucleons and mesons) in the medium can be obtained in the
relativistic mean field approach (RMFA) with the VF effects by
calculating the loop-diagrams in Fig. 1. We first come to the
self-consistent calculations of in-medium meson masses. The obtained
in-medium meson masses are presented in Fig. 2. It is obviously seen
from Eqs. (16), (17) and (20), (21) that the in-medium mesom masses
are related to the asymmetry parameter $\alpha$. The results in Fig.
2 are for both cases of $\alpha$=0.0 (symmetric matter) given by
VF$\rho$ and $\alpha$=1.0 (asymmetric matter) given by
VF$\rho\delta$ models, respectively.

\vspace{3cm}

\begin{figure}[hbtp]
\begin{center}
\includegraphics[width=4in]{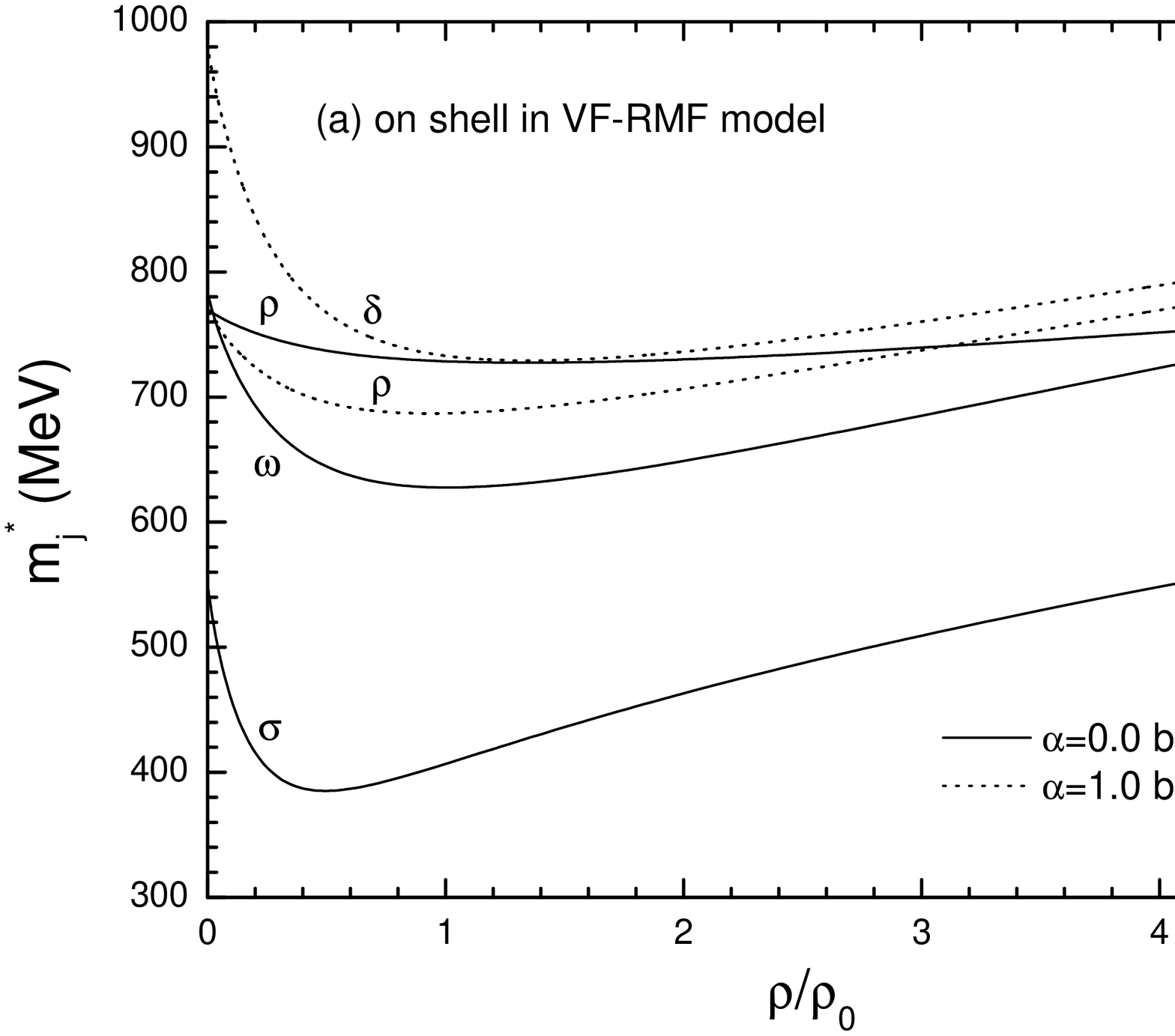}
\includegraphics[width=4in]{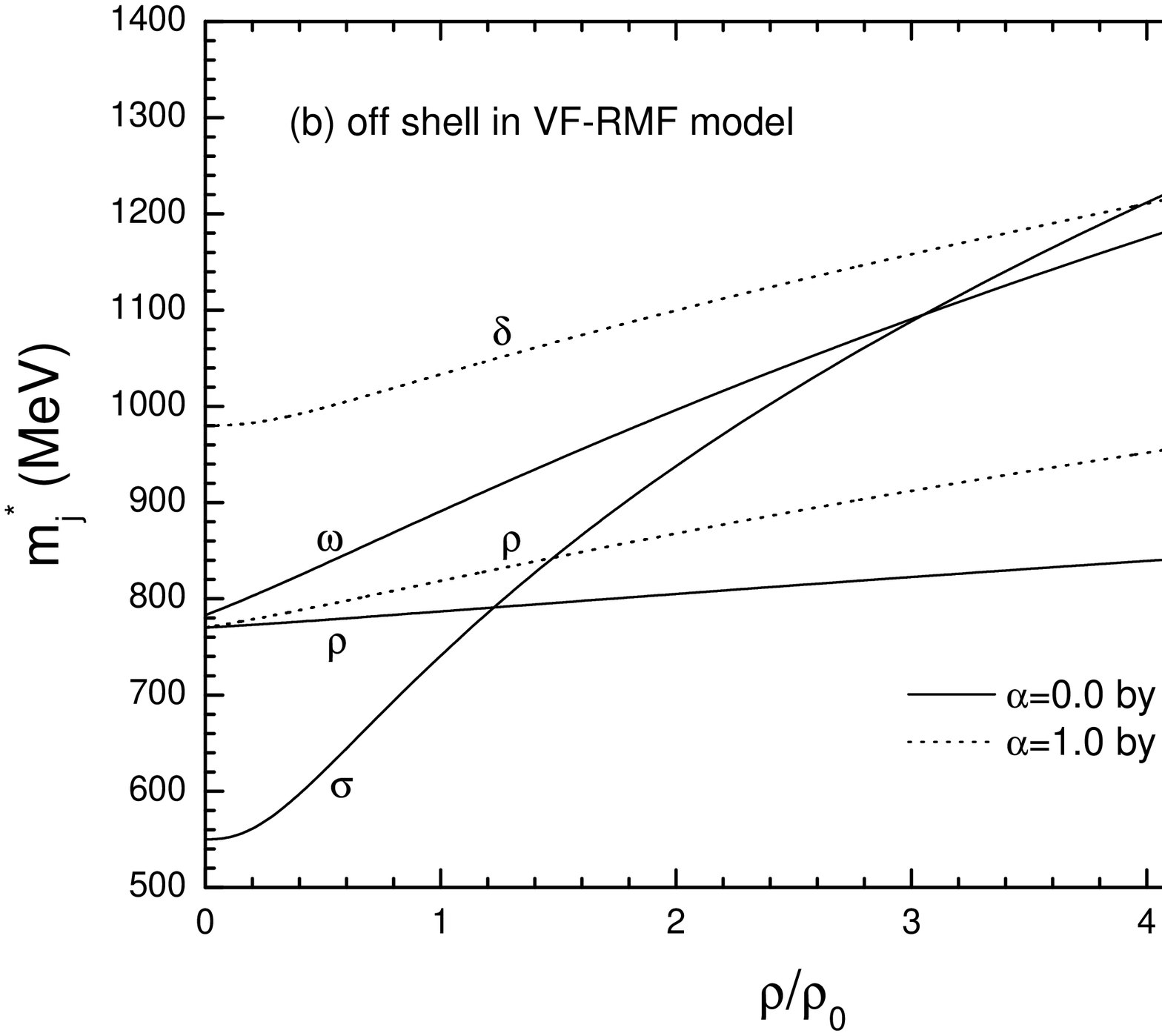}
\vglue -3.5cm \caption{Meson effective masses as a function of the
baryon density in the VF-RMF model.}
\end{center}
\end{figure}

Fig. 2(a) shows the in-medium meson effective masses (on-shell:
${\vec q}=0,~q^{0}=m_{j}^{\star},~j=\sigma,~\omega,~\rho,~\delta$)
as a function of the baryon density in the VF-RMF model.
 The in-medium modification to the masses of $\sigma$, $\omega$, and
$\rho$ mesons have been studied in other theoretical models
\cite{hatsuda,brown,sarkar}. The in-medium effective mass decrease
at the normal density is $\sim$18\% for $\rho$ and $\omega$ mesons
in the model based on QCD sum rule \cite{hatsuda}. The mass decrease
is $\sim$20\% for $\omega$ and $\rho$ mesons at the normal density
according to Brown-Rho (BR) scaling law \cite{brown}. In our model,
the decreases of in-medium meson effective  masses at the normal
density are $\sim$25\% for $\sigma$, $\sim$20\% for $\omega$, and
only $\sim$5\% for $\rho$ mesons in the symmetric VF$\rho$ case,
respectively. In the VF$\rho\delta$ case, the decreases of $\rho$
and $\delta$ mesons are 11$\sim$13\% and 25$\sim$27\% at the normal
density, respectively. Most experiments and theoretical approaches
indicated a decrease of the in-medium meson effective masses around
the normal density comparing with the masses at zero density
\cite{hatsuda,brown,sarkar,ozawa,trnka,krusche}. However, in the
latest experiment \cite{nasseripour}, no significant mass shift for
the $\rho$ meson with momenta ranging from 0.8 to 3.0 GeV was
observed. Up to now, the experimental results have not yet
converged, and more work is needed to obtain consistent
understanding of the in-medium behavior of vector mesons
\cite{hayano}. In general, the medium modifications to the masses of
mesons are momentum dependent \cite{post}. We note that the on-shell
meson effective masses are obtained for the mesons at rest in our
model. This is not in the momentum range of the CLAS experiment
\cite{nasseripour}. Until now, there is no experimental measurement
about the in-medium modification to the mass of the $\delta$ meson.
Our model indicates a significant decrease of the $\delta$ meson
effective mass around the normal density. We note that the effective
meson masses become to increase at high density regions in the
VF-RMF model. Unfortunately, the high density regions are beyond the
reach of current experiments. It will be very interesting to test
our prediction in the future experiments.

Fig. 2(b) shows the in-medium meson effective masses (off-shell:
$q^{\mu}=0$), which are used for the calculations of the nuclear
EOS, as a function of the baryon density in the VF-RMF model. The
off-shell meson masses are different from the on-shell meson masses
because the four momenta carried by the meson propagators in these
two situations are different. Since the meson propagators in the
nucleon self-energy are computed at zero four momentum transfer (see
Fig. \ref{fig01}(a)), we have to use the off-shell meson masses in
the tadpole loop calculation for self-consistency.

\vspace{3cm}

\begin{figure}[htbp]
\begin{center}
\includegraphics[width=4in]{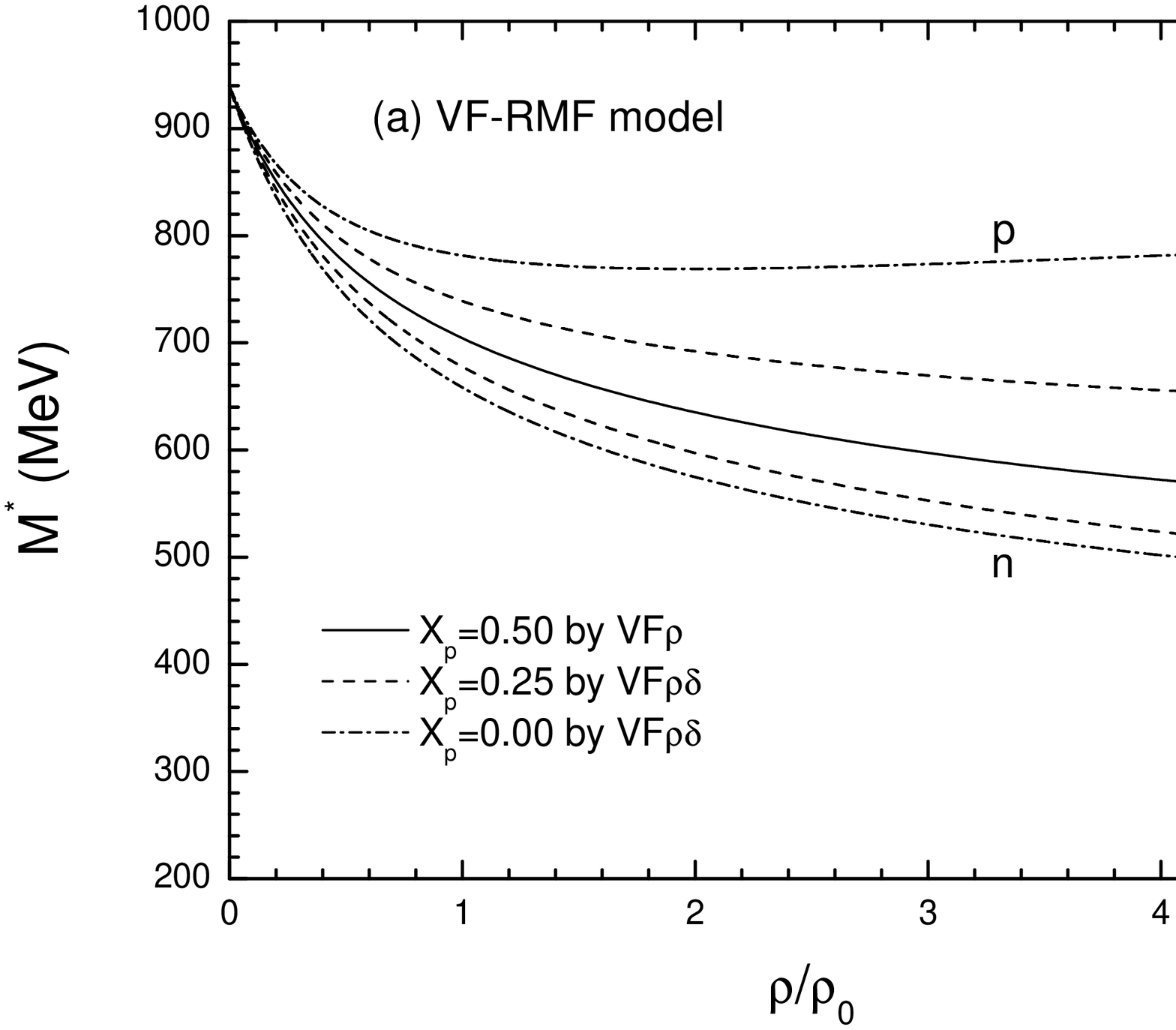}
\includegraphics[width=4in]{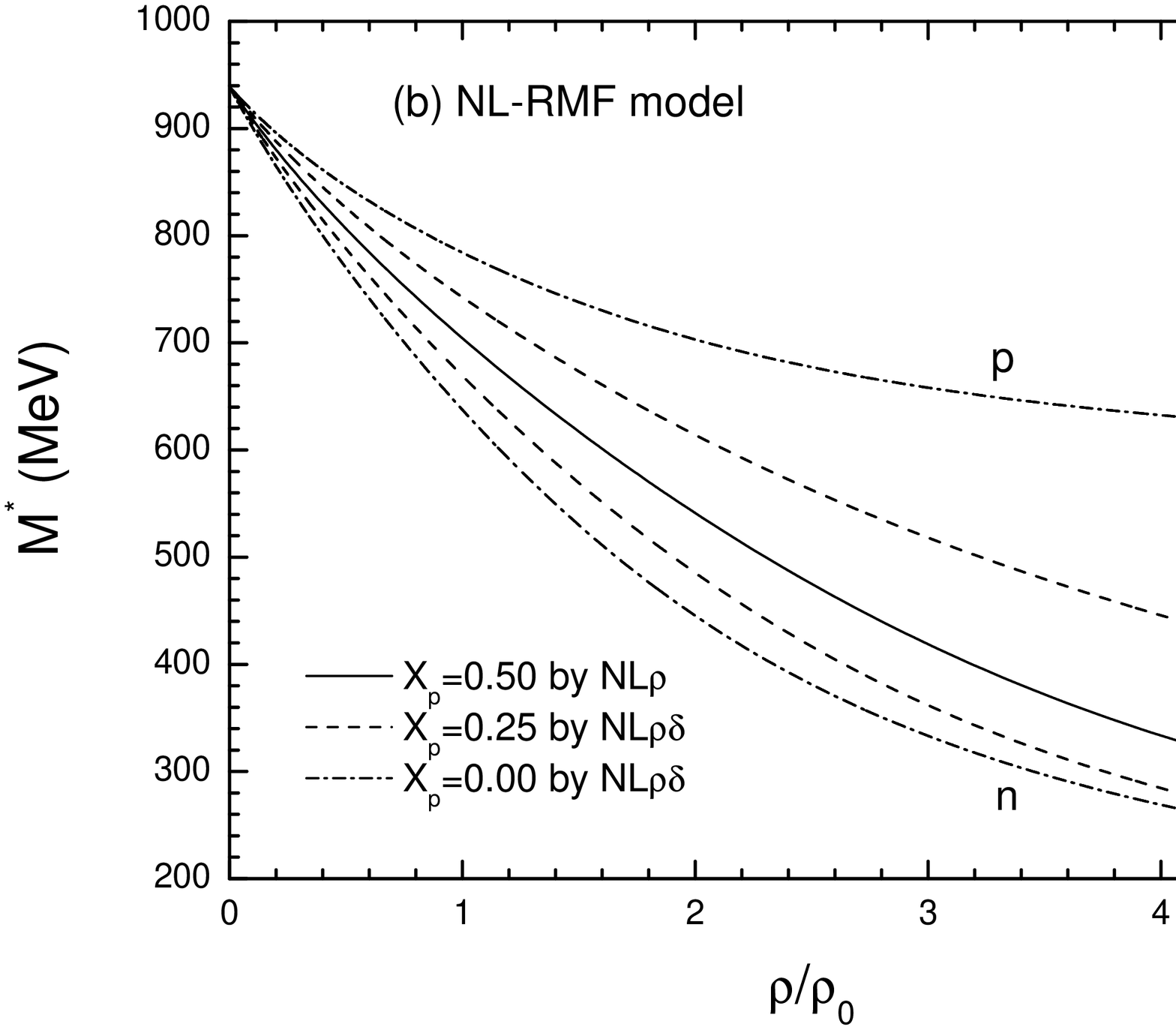}
\vglue -3.5cm \caption{Nucleon effective masses as a function of the
baryon density in different models. The upper (lower) dashed and
dash-dotted lines correspond to the masses of the proton (neutron).
(a) VF-RMF model. (b) NL-RMF model.}
\end{center}
\end{figure}

The nucleon effective masses play an important role in the
calculations of the EOS (see Eqs. (25) and (26)). We calculate the
loop-diagram corrections to the self-energies of nucleons in medium.
The nucleon effective masses without the $\delta$ meson can be
calculated from Eq. (7). One can see from Eqs. (12) and (13) that
the presence of the $\delta$ meson leads to proton and neutron
effective mass splitting. We present the baryon density dependence
of proton and neutron effective masses for different proton
fractions in the two models for a comparison in Fig. 3. The solid
lines in Fig. 3 are the nucleon effective mass for symmetric matter
($X_{p}$=0.5).

Fig. 3(a) shows that the proton and neutron effective masses given
by the VF$\rho\delta$ model decrease slowly with the increase of the
baryon density, at variance with the NL$\rho\delta$ model that
presents a much faster decrease (Fig. 3(b)). This main difference
between the VF$\rho\delta$ and the NL$\rho\delta$ models actually
comes from the in-medium meson masses (see Eqs. (22) and (23)). The
in-medium meson masses (off-shell) increase with the increase of the
baryon density in the VF-RMF model (see Fig. 2(b)). So it is natural
that the VF effects lead to a slow decrease of the nucleon effective
masses as the baryon density increases.

\vspace{-2.5cm}

\begin{figure}[htbp]
\includegraphics[scale=0.42]{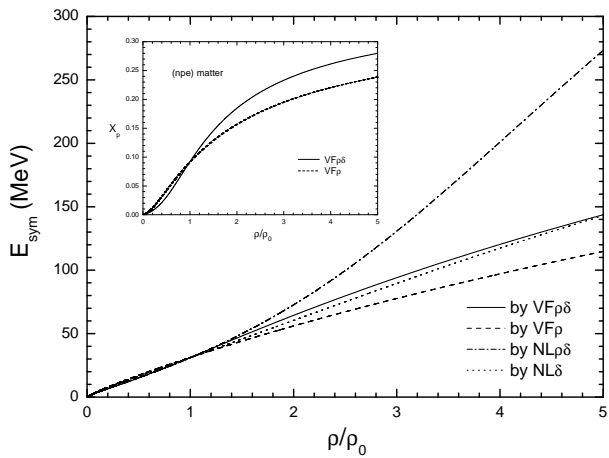}
\vglue -3.0cm \caption{The symmetry energy as a function of the
baryon density in different models. The inset is the corresponding
proton fraction.}
\end{figure}

The density dependence of the symmetry energy for the VF-RMF and
NL-RMF models is presented in Fig. 4. We see a similar behavior of
$E_{sym}$ at saturation density for both the two models. With
 the increase of the baryon density, the symmetry energy given by the VF-RMF
model increases slowly comparing with that given by the NL-RMF
model.
 From Fig. 4 we see that the symmetry energy with the $\delta$ meson is stiffer than that without
the $\delta$ meson for both the VF-RMF and the NL-RMF cases. The
symmetry energy in the VF$\rho\delta$ case is softer than that in
the NL$\rho\delta$ case. This is due to the VF effects. The presence
of the $\delta$ meson affects the symmetry energy and consequently
the EOS of asymmetric nuclear matter.

\vspace{2cm}

\begin{figure}[htbp]
\includegraphics[scale=0.35]{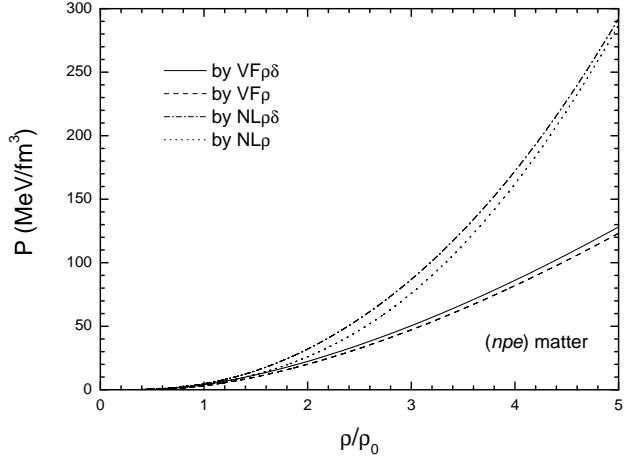}
\vglue -3.5cm \caption{Equation of state for ($npe$) matter.}
\end{figure}
\vspace{-2cm}

The $\beta$-equilibrium matter is relevant to the composition of the
neutron stars. The EOS, pressure vs density, for ($npe$) matter in
the VF-RMF and the NL-RMF models is presented together in Fig. 5 for
a comparison. We see that the EOS in the VF-RMF is lower. Due to the
VF effects, the EOS of asymmetric matter becomes softer.

In the present work, only two pictures for the neutron star
composition are considered: pure neutron and $\beta$-equilibrium
matter, $i.e.$ without strangeness bearing baryons and deconfined
quarks (see Refs. \cite{lattimer,maieron}). Furthermore, we limit
the constituents to be neutrons, protons, and electrons in the
latter case. In fact, in $\beta$-equilibrium matter, nucleons and
electrons indeed dominate at low temperature.

The structure of neutron stars can be calculated by solving TOV
equations. The correlation between the neutron star mass and the
corresponding radius for the pure neutron and the
$\beta$-equilibrium ($npe$) matter by the VF-RMF model are shown in
Fig. 6. The obtained maxium mass, corresponding radius and central
density are reported in Table III. We see from Fig. 6 and Table III
that the VF-RMF model leads to the decrease of the neutron star
masses
 for both the pure neutron and the ($npe$) matter.
 However, the NL-RMF model leads to heavier neutron stars (see Refs. \cite{liu05,liu0702}).
 This is mainly because the EOS of asymmetric matter
 becomes softer since the symmetry energy becomes softer in the VF-RMF model.

\vspace{3.0cm}

\begin{figure}[htbp]
\includegraphics[scale=0.35]{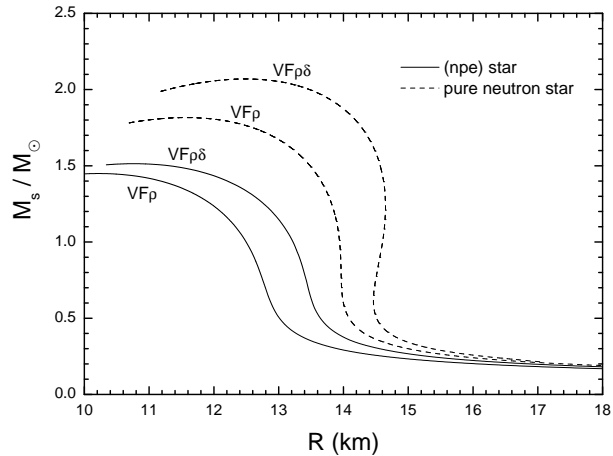}
\vglue -3.5cm \caption{The mass of the neutron star as a function of
the radius of the neutron star.}
\end{figure}

\begin{table}
\begin{center}
{{\large \bf Table III.}~~The maximum mass, the corresponding radius
and the central density of the neutron star in the VF-RMF model.}
\par
\vspace{0.5cm}
\noindent
\begin{tabular}{c|c|c|c} \hline
$   $   &$Model$  &\multicolumn{2}{c}{$VF-RMF$}   \\ \hline
$neutron~star$ &$properties$ &$VF\rho$ &$VF\rho\delta$  \\ \hline
$pure ~neutron$ &$M_{S}/M_{\bigodot}$
                                      &1.82   &2.07      \\ \cline {2-4}
                &$R (km)$             &11.57  &12.49     \\ \cline {2-4}
                &$\rho_c/\rho_0$      &6.41   &5.48      \\ \hline
$(npe)~matter$ &$M_{S}/M_{\bigodot}$  &1.45   &1.51      \\\cline{2-4}
                &$R (km)$             &10.22  &10.77     \\ \cline {2-4}
                &$\rho_c/\rho_0$      &8.98   &8.20      \\ \hline
\end{tabular}
\end{center}
\end{table}

We note that the coupling constant $f_{j}=g_{j}^{2}/m_{j}^{2}$ in the NL-RMF model
is a constant fixed by the saturation properties of nuclear matter \cite{liu02,liu05}.
In order to make a comparison, we define $F_{j}^{\star}=g_{j}^{2}/m_{j}^{\star 2}(\rho)$
($j$=$\sigma$, $\omega$, $\rho$, $\delta$) in our model, here $m_{j}^{\star}$
is the off-shell meson mass.
We are interested in the VF effects on the coupling constants. It is
well known that the coupling constants in the DDRH model are density
dependent (see \cite{liu0702}), whereas the meson masses are constant.
 In the present model, the coupling constants are constant, but the meson
masses are density dependent. In order to distinguish,
 we define $F_{j}=g_{j}^{\star 2}(\rho)/m_{j}^{2}~(j=\sigma,~\omega,~\rho,~g_{j}^{\star}$ is the coupling constant) for the
DDRH$\rho$ case. We present a comparison between $F_{j}^{\star}$ and
$F_{j}$ in Fig. 7. We see from Fig. 7 that $F_{j}^{\star}$ decreases
with the increase of the baryon density for both $\sigma$ and
$\omega$ mesons, which are due to the increase of the off-shell
meson masses in high density regions. It shows that the variance
trends of $F_{j}^{\star}$ and $F_{j}$ are roughly the same for
$\sigma$, $\omega$, and $\rho$ mesons. If we attribute the density
dependence of the coupling constants in the DDRH model to the VF
effects in our model, the density dependence of $F_{j}^{\star}$ and
$F_{j}$ should have the same physical origin.

\vspace{2.5cm}

\begin{figure}[hbtp]
\includegraphics[scale=0.35]{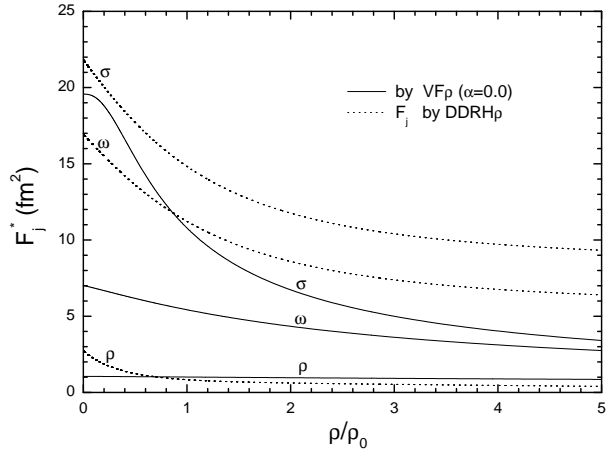}
\vglue -3.5cm \caption{A comparison between the ratio
$F_{j}^{\star}$ and $F_{j}$.}
\end{figure}

\section*{IV. Summary}

 The VF corrections are investigated
 in the framework of the RMF approximation
 by using the relativistic Lagrangian density with the $\delta$ field in this work.
By taking into account the loop corrections to the self-energies of
the in-medium nucleons and mesons,
 the VF effects are naturally introduced into the RMF model.
In order to make a comparison with the NL-RMF model, the same
saturation properties of nuclear matter are used to determine the
parameters of the VF model. We calculate the contributions from
self-energy diagrams to the masses of the nucleons and mesons. The
effective masses of the nucleons and mesons, especially the $\rho$
and $\delta$ mesons, in the nuclear medium are obtained. We find
that the nucleon effective masses decrease more slowly with the
increase of the baryon density comparing with the NL-RMF case.
 We also find that the dependence of the off-shell meson masses on the density
 in the medium is different from that of the on shell meson
masses (see Fig. 2). The effects of the in-medium hadron masses on
the nuclear EOS and the properties of the neutron stars are studied.
The VF effects lead to softness of the symmetry energy and the EOS
of asymmetric matter. Due to such softness, the neutron star masses
are reduced quite a lot. This indicates that the VF effects on the
neutron stars are important. We see from Fig. 2(b) that the
off-shell in-medium meson masses increase with the increase of the
baryon density. The off-shell in-medium meson masses can be used to
calculate $F_{j}^{\star}$. The variance trends  of  $F_{j}^{\star}$
for $\sigma$, $\omega$, and $\rho$ mesons are roughly consistent
with $F_{j}$ in the DDRH$\rho$ case \cite{liu0702}. The density
dependence behavior of the off-shell in-medium meson masses is very
interesting indeed.

In the present work, we work in the Hartree approximation in which
we only consider the dominant contributions from tadpole diagrams to
the nucleon self-energy when we investigate the VF effects. In fact,
the exchange diagram contributions can only provide small
corrections to the EOS in the RMF approach at high densities
\cite{brian}. It is well known that the symmetries of the infinite
nuclear matter system can simplify the mean-field Lagrangian
considerably. As pointed out in Ref. \cite{brian}, when
translational and rotational invariance of the nuclear matter is
taken into account, the expectation values of all three-vector
fields must vanish. Therefore, the expectation value of
$\vec{G}_{\mu\nu}$ in Eq. (1) is zero and nonlinear $\rho$ meson
interactions (three or four $\rho$ vertices) do not appear.
Furthermore, since we only take into account the tadpole diagrams,
only the neutral iso-vector meson ($\rho^{0}$) is involved in the
nucleon self-energy diagrams even in the asymmetric nuclear matter
($\rho^{\pm}$ could contribute to the exchange diagrams which are
ignored in the Hartree approximation).

It is suggested that the tensor coupling to the nucleon should be
taken into account for the study of the vector mesons in nuclear
medium \cite{machl}. In the present work, since we focus our
attention on the VF effects on the properties of neutron stars, we
do not include the tensor coupling effects in the calculations for
the effective masses of the vector mesons. This will be studied in
the future work, especially for the study of the $\rho$ meson masses
in the medium. Furthermore, the future study of the VF effects on
the meson-nucleon coupling constants will be carried out. In
addition, more careful study of the high density behavior of the
meson-nucleon effective couplings, especially $g_{\delta}$, is
important and attractive.

\begin{acknowledgments}
  This project is supported by the National Natural
  Science Foundation of China (Project Nos. 10675022 and 10875160), the Key
  Project of Chinese Ministry of Education (Project No. 106024), and
  the Special Grants from Beijing Normal University.
 \end{acknowledgments}

\end{document}